%% file: 2019-ddos-call-paper.tex
\documentclass[acmsmall]{acmart}

\setcopyright{none}
\renewcommand\footnotetextcopyrightpermission[1]{}

\usepackage{balance}

\usepackage[inline]{enumitem}

\usepackage{color,soul}

\begin{document}

\title{20 Years of DDoS: a Call to Action}
%%
%% LZ: Maybe not "Happy"?  how about "20 years of DDoS: a call to action"?
%%
\author{Eric Osterweil}
\affiliation{
	\institution{George Mason University}
}
\email{eoster@gmu.edu}

\author{Angelos Stavrou}
\affiliation{
	\institution{George Mason University}
}
\email{astavrou@gmu.edu}

\author{Lixia Zhang}
\affiliation{
	\institution{UCLA}
}
\email{lixia@cs.ucla.edu}

\input{abs}





\maketitle

\input{intro.tex}

\input{bg.tex}
\input{sou.tex}

\input{arch-sou.tex}
\input{vol-sou.tex}

\input{econ-sou.tex}
\input{scrubbing.tex}

\input{remediation.tex}

\input{disc.tex}

\input{acks.tex}

{ \balance
{
	\bibliographystyle{ACM-Reference-Format}
	\bibliography{ddos.bib}
}
}

\end{document}

%% file: abs.tex
\begin{abstract}

Botnet Distributed Denial of Service (DDoS) attacks are now 20 years old; what has changed in that time?  
Their disruptive presence, their volume, distribution across the globe, and the relative ease of launching them have 
all been trending in favor of attackers. 
Our increases in network capacity and our architectural design principles are making our online world richer, but are favoring attackers at least as much as Internet services. 
The DDoS mitigation techniques have been evolving but they are losing ground to the increasing sophistication and diversification of the attacks that have moved from the network to the application level, and we are operationally falling behind attackers.
It is time to ask fundamental questions: are there core design issues in our network \emph{architecture} that fundamentally enable DDoS attacks?  How can our network
\emph{infrastructure} be enhanced to address the principles that enable the DDoS problem?  How can we \emph{incentivize} the development and deployment of the necessary changes?
In this article, we want to sound an alarm and issue a call to action to the research community.  
We propose that basic research and principled analyses
are badly needed, because the status quo does not paint a pretty picture for the future.

\end{abstract}

%% file: intro.tex
\section{Introduction}
\label{sec:intro}

{\bf What has happened in the last 20 years:} \newline
In 1999, a set of compromised computers, called Trin00~\cite{trin00}, took down a network at the University of Minnesota; and 
with this first documented case, 
botnet volumetric Distributed Denial of Service (DDoS) attacks were undeniably born.  
While earlier attacks against infrastructure exist in anecdotes and recollections, it is with this documented case that we can archivally establish the lower 
bound of 20 years.  Many changes, enhancements, and evolutions 
to our mitigation technologies
%%
%% LZ: of what? you mean Many changes, enhancements, and evolutions of the mitigation technologies?
%% EO: Fiixed, thanks
%%
have happened since then, but are we demonstrably better off today (now 
% at least 
20 years later)?
Trin00 used hundreds (and actually may have been composed of thousands) of compromised machines (``bots'').
Today, conventional bot-network (``botnet'') sizes have been seen in the millions.  
By today's standards, Trin00 may not sound like a large botnet.  
However, size is not all that matters.  In some cases, smaller (well-provisioned) botnets have hit harder than any before (e.g. the Mirai botnet attacks~\cite{mirai-botnet}).
Attacks like those from Mirai illustrate that the \emph{size} of a botnet is not the sole determining factor in the damage it can do.
Another, perhaps archival, lesson is that historical attack sizes are relative, and raw numbers alone don't tell the tale.
Moore's law and bandwidth increases make comparing attack volumes (bits-per-second, bps) from the past to today an apples to oranges comparison.
Gigabit attacks in 2000 were considered staggering, but only because they rivaled provisioned capacity of services and carriers of the time.
An unfortunate state of affairs is (and has been for 
% at least 
the last 20 years) that it is easier to gain attack capacity (i.e. compromised hosts and aggregate 
attack bandwidth) 
than it is to gain defensive capacity.
Moreover, during the last 20
% + 
years, our remediation strategies have not fundamentally evolved as attacks have been swelling in size and complexity.  
The state-of-the art in the DDoS defense industry still centralizes our defenses (in ``scrubbing'' centers) against growing distributed attacks.
In short, what was true then is still true now: DDoS is an asymmetric threat with impedance mismatch between attackers and defenders,
which strongly indicates the need to reexamine the principles that underlie the problem-space.

\noindent
{\bf Victims of our architecture's success:} \newline
Internet protocols have long been designed to abstract decisions and operations away from each other to foster heterogeneity, scalability, encapsulated functions, and more.
At the same time, DDoS attacks have evolved to using Techniques, Tactics, and Procedures (TTPs) at and above the network layers (at the application-level).
This has, as a consequence of our layered abstraction model, hidden DDoS semantics from the transport and network layers.
As defenders against DDoS attacks, our fundamental challenge is the onus to tear apart attack traffic from legitimate traffic, where the distinction is often only visible at the application layer, and conventionally encrypted there.
A recent operational report of large-scale measurements stated ``SSL [sic] is \underline{majority} of traffic in [North America] by 2019''~\cite{labovitz-apricot19}.
The necessary computational complexity, the volume of traffic, and the growing use of encryption often render common operational network tools ineffective in defending against attacks.
% GH citation
% What's more, some operational observations infer that the deployment of necessary security protections are often limited by ``The \emph{costs} $\dots$ not directly [being] borne by the potential beneficiaries of deploying the solution''~\cite{huston-apricot19}.
What's more, some operational observations infer that the deployment of necessary security protections has been limited ``Where there is no clear 
early adopter advantage''~\cite{huston-apricot19}.

In facing the distributed threat of DDoS, our distributed network \emph{should} be our greatest countermeasure, but how without application-level inspection of traffic?  
For example, how should a network-level management tool determine which of a stream of DNS queries are real and which are participating in a reflector attack~\cite{Guo:2006:SDP:1153919.1154048}? 
Or, which NTP command is legitimate and which is part of an attack~\cite{Czyz:2014:TPG:2663716.2663717}? 
Or, is a {\tt memcached} query from a real application or part of an attack~\cite{baiaanalysis}?
Or, which HTTP client is trying to keep a needed connection alive and which is starving the server for resources~\cite{6623707}.
Today, it is hard for either the network-layer or the transport-layer to mount an effective defense because they do not have policy semantics that can encode diverse application-level nuances.
DDoS mitigation is necessarily done using Deep Packet Inspection (DPI) and, therefore, only after centralizing distributed attack traffic.
This impedance mismatch, distributed attack versus centralized mitigation, is further complicated when application payloads are ``embedded'' (e.g. encrypted) and require multiple layers of complicated and expensive ``decoding.'' 
For example, performing DPI on an HTTPS flow requires decryption of the flow, but \emph{that} requires escrow of the end-site's TLS private key 
(to terminate and inspect the embedded flow).
Internet protocol layering, coupled with the end-to-end principle, has made it difficult to shut down the DDoS floodgate because bad traffic cannot easily be filtered out at the network layer alone.
In short, we have been losing the DDoS war, and it is time to ask \emph{why} and investigate the fundamentals of this problem-space.

% , and we need the research community to turn the tide.
In this article, we want to sound an alarm that we must take corrective action.
We posit that the research community is ideally 
suited to formulate and investigate \emph{fundamental} questions about how we got here.
For example,
are there foundational issues in our network \emph{architecture} that fundamentally enable DDoS attacks?  How can our network
\emph{infrastructure} be enhanced to address the principles that enable the DDoS problem?  In order to reap the gains of our work, how can we \emph{incentivize} the 
development and deployment of necessary changes?
%
% The research community is, perhaps, uniquely qualified and posiitoned to take a principled look at what factors are fundamentally enabling DDoS attacks.
To move our defensive posture forward we need to take a fresh look at the problem and consider fresh approaches.

This article is structured as follows:
to illustrate how dire our situation is, we provide a brief background, then propose a few ways to categorize 
the ``state of the union'' of DDoS attacks: architecturally, volumetrically, and economically.  
We then take a closer look at the fundamental nature and state of DDoS attacks, today.
From there, we discuss modern mitigation techniques in the DDoS defense industry (e.g. ``scrubbing''), before concluding with a discussion.

%% file: bg.tex
\section{DDoS Background}\label{sec:bg}

There is not one type of DDoS, there are a variety.  Some are called ``low-and-slow,'' which starve servers of resources, and can be hard to detect.
Some are volumetric, which send overwhelming amounts of traffic that congest network links and overload servers and are hard to stop even though they are detectable by nature.
There is a large body of literature that broadly proposes DDoS taxonomies~\cite{mirkovic2004taxonomy,asosheh2008comprehensive,ddos-protection-center}.
Much of the prior work has included more than just DDoS attacks, themselves, but also categorized malcode ecosystem aspects
like implementation details, malcode behavioral differences, 
the infection vectors, and also the attack TTPs involved in DDoS attacks.
% To highlight the status of our war on DDoS, here,
% we focus primarily on the natures of DDoS attacks, and for clear illustration, we further focus briefly (below) on one specific common type: the clear-and-present threat posed by 
% volumetric DDoS.\footnote{We note that
% singling our volumetric attacks is not meant to undercut the importance of other types of DDoS or to suggest our observations are limited to them, but simply to present 
% some specifics of one common type of attack.}
% 
In order to highlight the status of our war on DDoS, we focus on addressing operational aspects of DDoS attacks: their natures, their traffic, and
detecting and remediating them.
Of the many types of DDoS attacks and TTPs, we briefly describe a couple common generalized examples:
volumetric and resource starvation.\footnote{This is not meant to be presented as a comprehensive list, a taxonomy, or to undercut the importance of other types of DDoS, 
or even to suggest our observations are limited to these.}

\subsection{Volumetric DDoS, today}
Large volumes of DDoS attack traffic (sometimes called ``packet love''), generally require service providers to invest in very expensive infrastructure and 
network bandwidth (capacity).
In volumetric DDoS, the largest recorded DDoS attacks have all been stateless and have capitalized on
the ability to spoof (or ``lie about'') the source addresses of their attack traffic.  
As just a couple of examples:
in 2016, the first publicity around a terabit attack came from an attack on 
krebsonline.com~\cite{krebs2016krebsonsecurity}) and it was able to reach this volume through its use of source address spoofing in a reflector attack.
Not long after that in 2016, a larger attack on Dyn~\cite{mansfield2016ddos} surpassed this volume, again using source address spoofing. 
% TTPs of today’s volumetric DDoS attacks
% Spoofed vs. non-spoofed source-addresses
In short, the TTPs of the largest DDoS attacks seen today rely heavily on being able to spoof addresses (even if that is just to leverage an online service for an
amplification factor).
% Shift of usage from stateful to stateless transport protocol

While this is not a new type of DDoS, the increasingly relative ease of acquiring the disruptive power of large volume attack sources has elevated the appeal of stateless
DDoS attacks to adversaries.
% UDP (DNS, SNMP, NTP, etc.)
% Key to this type of attack is the use of datagrams with spoofed source addresses.  These may be Domain Name System (DNS) queries, Simple Network Management Protocol (SNMP) queries,
Spoofed datagrams may be Domain Name System (DNS) queries, Simple Network Management Protocol (SNMP) queries,
Network Time Protocol (NTP) queries, {\tt memcached} queries, or others.
% Some are direct attacks from well provisioned sources
% SYN floods (with data) (MIRAI, XOR, etc.)
Some stateless attacks may also be spoofed control traffic for TCP connections (not reflected traffic).  
For example, TCP SYN packets that have spoofed source addresses and large data payloads~\cite{xor-attacks-csoonline} (even though SYN 
packets are not permitted to carry data on setup~\cite{postel1981rfc}).

%% Observe fundmantal weakness, long existance, growing ease of use, lack of fundmanetal remediation techniques: unmaintainable asymmetry 
% Considering the destructive power of these attacks, how long we have had to endure them, and the rate at which they have grown, a timely question is: 
% is there something fundamental that is enabling them?  This is a deceptively complicated question, and we maintain it is a open research challenge.  
% Nevertheless, it is worth noting that it remains relatively easy for malicious users to acquire bots for a botnet, there is a clear asymetry in remediating 
% distributed attacks in a centraliized way, 

\subsection{Resource Exhaustion Attacks}

While many headlines focus on the largest DDoS attacks seen, and in recent years, those have all tended to be stateless, there continue to be many instances of
attacks that leverage protocol aspects in the Secure Socket Layer (SSL), Transport Layer Security (TLS), or even at the HTTP layer (which might be using HTTPS).
These attacks can be crippling without a DDoS defense system, but they do not result in headlines as often as their volumetric counterparts.
% Stateful (i.e. TCP-based, not just SYN-flood) attacks have become less common, and have not approached the scale of stateless attacks
While these attacks don't approach the volumetric scale of reflector attacks, resource exhaustion attacks allow attackers to bring Internet services down with far fewer resources.

Perhaps the earliest known resource exhaustion attacks were those that abused the Transmission Control Protocol (TCP), itself: SYN-flood attacks.  
Reported evidence~\cite{meme-2.12}
suggested that these Denial of Service (DoS) attacks did not appear to be distributed, but this nevertheless illustrates that this type of attack has been used in the 
wild 
since at least 1996~\cite{center1996cert}.  
Attacks like these (which used source spoofing and may or may not be distributed) were initially intended to exhaust servers' resources, and were neither 
volumetric, nor stealthy (low-and-slow).\footnote{Though, we note that many recent attack TTPs incorporate this technique in distributed attacks (DDoS).}

% Slowlaris
One of the early examples of low-and-slow attacks was an attack called Slowloris~\cite{hansen2009slowloris} in which a relatively small number of stateful HTTP 
queries would hold connections open on webservers and thereby exhaust their ability to answer other (legitimate) clients.
% TLS
Other exhaustion attacks exploit TLS' cryptographic key negotiation~\cite{castelluccia2006improving}.  In these types of attacks, the raw numbers of attacking
clients and traffic are not as spectacular as volumetric DDoS, but (perhaps more troubling) is the fact that their \emph{detection and remediation} more clearly requires additional state information \emph{above the network layer}.

\subsection{Detection vs. Remediation:}

An important distinction in the DDoS war is the difference between \emph{detection} and \emph{remediation}.  The techniques to detect a DDoS often are very different than remediating
it.  Moreover, detection and remediation are not always addressed in the same places in the network, or even by the same service provider.  One common mode of operation is for
an online service to detect that it is under DDoS attack (of some kind), and to then engage a DDoS mitigation provider.  We discuss this in more detail in Section~\ref{sec:scrubbing}, after we first discuss the DDoS State of the Union.

%% file: sou.tex
\section{State of the Union}\label{sec:sou}

\noindent {\bf Assessment: are we winning, or losing?} 
There are many ways one could evaluate the state of the union with respect to DDoS.  Here, we use three example perspectives to categorize ways in which our status could be evaluated:
% We look at three measures of how successful we are in the DDoS wars:
\begin{enumerate*}
\item{architectural}
\item{volumetric}
\item{economic.}
\end{enumerate*}
These are not meant to be \emph{the} canonical set, a complete set, or to be a formal framework.  Rather, these are simply used to help illustrate aspects of the DDoS defense ecosystem.

%% file: arch-sou.tex
\subsection{Architectural State of the Union}\label{sec:arch-sou}

Technologically, not much has \emph{fundamentally} changed for defenders, but \emph{a lot} has changed for the attackers.
Our reliance on DPI for detecting and remediating attack traffic has resulted in increasing dependence on keeping our defenses in centralized  approaches (relative to DDoS sources).
For example, with reflector attacks leveraging application-level semantics, and the increased use of Transport Layer Security (TLS), terminating and interpreting traffic has necessitated 
backhauling traffic to DPI, or ``scrubbing,'' centers.
% We discuss this in more detail in Section~\ref{sec:scrubbing}.
There, the tools used by defenders have incrementally evolved, but the fundamentals of our approaches have not.
What's more, 
this has framed an architectural asymmetry: large volumes of attack traffic (more sources with increasingly better provisioned networks) vs. central remediation.
This is particularly worrisome in the face of the increased complexity of web applications and their increased use of encryption.
These often demand that a remediation engine act as an end-point in a network flow.
The architecture of today's DDoS mitigation techniques is centered around machines (or network appliances) that inspect traffic, both generally and at the application-level.  
% There exist 
% many commercial services dedicated to selling mitigation services to other online providers~\cite{akamai-prolexic, neustar-ddos, cloudflare-ddos}.
% These companies sell mitigation services (Mitigation as a Service, MaaS) on well-provisioned networks, and then draw attack traffic across their scrubbers (DPI appliances).
As a result, our mitigation techniques are predicated on matching mitigation bandwidth to ever-growing aggregate \emph{distributed} attack volumes, 
and that is (at best) a band-aid solution.
This observation has echoed if different ways, and in different places for some time.
For example, in 2015, the Defense Advanced Research Projects Agency (DARPA) announced a call for Extreme DDoS Defense (XD3) that included a solicitation 
to ``\emph{[disperse]} cyber assets (physically and/or logically)''~\cite{darpa-xd3}.

% Blackholing the attack requires detection of the attack and attribution to sources which is not always easy for Application-borne attacks.
Some existing operational attempts towards \emph{dispersing} network-based remediation, such as BGP's FlowSpec \cite{RFC5575}, Remote Triggered Back-Holing (RTBH)~\cite{RFC5635}, etc. 
attempt to coordinate \emph{distributed} defenses 
by pushing remediation information to the network-layer.
However, without the necessary application-level expressiveness, this can unfortunately lead to collateral damage 
%For example, RTBH is often maligned because entire network prefixes are indicted, which can cause collateral damage 
to well-behaving sources contained within the same network prefix as attackers, non-attack traffic that is sourced from compromised devices, etc.,
and has arguably limited the appeal and adoption of these protocols and techniques.

%% file: vol-sou.tex
\subsection{Volumetric State of the Union}\label{sec:vol-sou}

Considering the volumetric state of the union (volumes of attack traffic vs. carriers/providers provisioned capacities) paints a similarly disconcerting picture.
Service Providers (SPs) buy transit in Gigabit per second links (Gbps) in multiple locations from multiple carriers.
Internet eXchange Points (IXPs) and carrier capacity are also often offered in Gbps.  Large carriers' global aggregate capacity may approach (and in some cases achieves) 
Terabits per second (Tbps), but this does not mean any given ingress point to a carrier's network is (itself) a Tbps link.
Generally, aggregate capacity in Tbps is summation of router/regional capacities (Gbps).
Consider, even routers' linecards only reach 100 Gbps, but aggregate attack traffic of the largest DDoS attacks is already over 1 Tbps.
In an aggregate view, a recent observation from operational measurements quotes that ``attacks [are] growing in size faster than network growth.''~\cite{labovitz-apricot19}
% We build inspection appliances (DDoS-defense/DPI boxes) and sell mitigation services (Mitigation as a Service) on well-provisioned networks, and then draw attack traffic across them
% This back-of-the-envelop analysis is underscored by operational measurements that report DDoS attack growth is already exceeding transit capacity growth~\cite{labovitz-apricot19}.
% For many service providers, the aggregate network capacity can often be measured in \emph{hundreds} of Gbps, \emph{globally.}
% It is certainly true that some of the very largest service providers have capacities in Tbps globally.

Moreover, the aggregate capacity matters far less than if this capacity exists near all attack sources.
Often, it is all but guaranteed that aggregate capacity is \emph{not} near attack sources, and it can often be topologically very far from attack sources.
``The Internet’s capacity attenuates the total throw weight a DDoS attack can generate; the farther a target is from components of a network, the less traffic that 
will make it across any congested links between the target and the attack source''~\cite{akamai-soti18}.  This can (and often does) result in service degradation and outages 
to \emph{other} Internet services, whose traffic shares congested routing infrastructure (i.e. collateral damage).

% If not near attack, backhaul attack traffic over transit links?
When attack sources are topologically far from mitigation, their traffic is backhauled across transit and peering infrastructure to scrubbing centers.
This has the effect of centralizing distributed attack traffic for mitigation, and draws terabits of attack traffic to (in some cases) gigabit scrubbing centers.
% Fundamentally lower capacity when centralizing
Even in the case of high-capacity scrubbing centers, the current state of affairs is that a distributed attack is necessarily remediated in a (relatively) 
more centralized mitigation infrastructure.

% Contrast: DDoS attacks are already bigger than many of our carriers’ capacities
The largest DDoS attacks that we have seen are already larger than the provisioned capacity of many (if not most) of
the large providers and carriers' capacities.
% In 2016, the fear was that ``one day'' DDoS could swell to 1Tbs~\cite{dhs-ddosd}
% In 2017, DDoS had reached 1Tbs
% In 2018, DDoS have exceeded 1Tbs
In 2016, the US Department of Homeland Security (DHS) started a program called DDoS Defense (DDoSD), whose starting position was, ``one day'' DDoS could swell to 
1 Tbps~\cite{dhs-ddosd}.
By 2017, the largest DDoS attacks had already reached that, and in 2018 DDoS attacks quantifiably exceeded that.

%% file: econ-sou.tex
\subsection{Economic State of the Union}\label{sec:econ-sou}

Using money as a canary-in-the-coalmine, service providers' outlay to protect against DDoS also paints a grim picture.
% If anything, things are getting worse, not better!
In 2000, DDoS attacks on Yahoo, eBay, and several other major Internet services led the news and raised alarms.
Now, almost 20 years later, protection rackets exist in gaming spheres.
Online gaming and gambling sites are frequently held hostage for ransom by DDoS threats~\cite{MANSFIELDDEVINE201513}, and sometimes attacks are launched simply 
in order to gain gaming advantages~\cite{10.1007/978-3-319-45719-2_17}.
More generally, today, all online services need DDoS protection, and companies expect to pay for for defensive protections against inevitable DDoS attacks.
The DDoS mitigation market was \$1.94 \emph{billion} in 2018, and is growing~\cite{ddos-market-size}.
What's more, there has also been a DDoS-for-hire (i.e. ``booter'') grey-market for roughly a decade~\cite{krebs-pp-booter, santanna2015booters}!

Internet services need protection, but there is no official remedy to DDoS.  So we must pay for help.
In a sense, we have privatized our police force (i.e. the DDoS mitigation marketplace).

%% file: scrubbing.tex
\section{DDoS Scrubbing}\label{sec:scrubbing}

Detecting an attack (versus other abnormal, but benign, traffic) can be critical, especially if a service provider intends to use a DDoS mitigation service to remediate attacks.
Under ``peace-time'' conditions, services are often provisioned to handle expected load and user behavior, periodic above-average ``bursts'' of traffic, slow TLS handshakes,
and other suboptimal client behaviors.  
% It is also common for service networks to be provisioned with large ``burst capacities'' (to service periods of above-normal traffic volumes).
Once an attack is detected, the subsequent remediation is very often done by applying application-level semantics to packets and/or packet flows.
These semantics allow remediation to inspect and ``scrub'' attack traffic off of legitimate traffic.

\subsection{Why scrubbing}
After DDoS traffic is ``scrubbed'' away, the remaining (``good'') traffic is then delivered to applications.
% Used to remediate both stateless and stateful attacks
% Common + vendor proprietary techniques inspect application-level attributes to determine “good” vs. “bad” traffic
Scrubbing uses techniques that range from measuring traffic heuristics to any number of vendor-specific techniques that assess the veracity of traffic and approaches 
that address
both stateless and stateful attack TTPs.
% At its most granular, scrubbing is done on a per-packet or per-flow basis, and not necessarily per source (i.e. at the network-layer).
Scrubbing, therefore, necessarily must have the ability to detect application-layer attacks and discern them from normal traffic (even if that normal traffic is just higher volume than
usual).
% Appliances: Arbor, Radware, A10, etc.
Some commercial solutions for on-premises mitigation appliances are Netscout's Arbor, Radware, and A10 Networks~\cite{arbor, radware, a10-ddos}.

Traffic sources (either remote networks or specific remote hosts) might be sending both proper application traffic and DDoS traffic.  For example, a single large home-access network
(under a single routed BGP prefix) might have well-functioning hosts transacting with a website, and separate compromised hosts (bots) sending DDoS traffic (possibly some hosts
sending both attack and non-attack traffic).

% E.g. is a DNS packet legit? Is an NTP query just trying to amplify attack traffic? Is a SYN actually trying to setup a TCP connection, etc.
This can make remediation difficult; for example, is a UDP packet a legitimate DNS query, or a legitimate NTP query, or is it a SYN packet that is \emph{actually} 
trying to setup a TCP connection, \emph{or} a legitimate TLS 1.3 0-RTT resumption?
In these types of situations, it can be very hard to separate attack traffic from non-attack traffic, solely at the network/transport layers.
% We need to keep up with TTPs by looking at the application-level semantics in traffic
Some attacks require multiple round trips with a source to distinguish, and
defenders are often very adverse to dropping legitimate traffic (false positives). 
In order to discern attack traffic from non-attack traffic, especially as TTPs continue to become more complex,
remediation often occurs by assessing application-level semantics of traffic.
% Scrubbing appliances and software inspects the transport and application-level semantics of traffic
This is precisely the remit of ``scrubbing'' appliances.  Their job is to ``scrub'' attacks out of, and forward on only, legitimate traffic.

% Regardless of which type of DDoS TTPs are the most prevalent at any given time, scrubbing facilities necessarily address as many as possible.
This unavoidable complexity is precisely why scrubbing is the industry's last line of defense.  While approaches like FlowSpec, RTBH, and the IETF's new working group
on DDoS Open Threat Signaling (dots)~\cite{ietf-dots} are all attempting to enhance the network/transport layer to aid in DDoS defense, scrubbing centers
catch all DDoS attacks that get through.\footnote{And, as of this writing, most DDoS attacks are mitigated in this way.}
% As scrubbing is a fallback (last resort), these solutions are used for both stateless and stateful attacks.

Because discerning the difference between a traffic burst and an attack can be difficult (and time is often of the essence), some Internet services
engage an ``always-on'' mitigation provider (sending all traffic through mitigation machinery, even during peace-time), so that detection and mitigation can both be handled together, with high confidence and low latency.
% traffic often must be detected and remediated from application traffic at the application layer

\subsection{State of the art: scrubbing centers}
%
% Centralized our defenses, against distributed attacks
% Victims route all of their traffic to well provisioned “scrubbing centers”
% DDoS scrubbing appliances and service centers are the primary mechanism by which DDoS attacks are miitigated, today.  
Whether scrubbing appliances are deployed
deeply in carrier networks' cores, or in service providers that draw traffic in, they represent a relatively centralized solution to DDoS' distributed threat.
% Services: Akamai, Neustar, Cloudflare, etc.
Sites that have a lot of bandwidth and specialized hardware and/or software are called scrubbing centers, and some companies offer these as a commercial service 
(Mitigation as a Service, MaaS)~\cite{akamai-prolexic,neustar-ddos,cloudflare-ddos}.
% Tbps networks, appliances and custom techniques, globally distributed, 24/7 SOCs, threat intelligence teams, 
While mitigation appliances are sold to SPs, and can be deployed in any network for self-protection, mitigation providers quantifiably offer more utility than trying to
detect and remediate DDoS on one's own.  Mitigation providers will have already invested in large transit and peering capacities (often in excess of 1 Tbps, in aggregate), 
will have deployed their infrastructure across the Internet at topologically diverse locations, likely have augmented mitigation appliances' technology with custom enhancements,
will have 24/7 Security Operations Centers (SOCs) monitoring traffic, and likely also have in house (or retained the services of) cybersecurity threat intelligence teams.  
%
% Ability to attribute attacks to specific source malcode (to discern TTPs, etc.)
% While this places a lot of onus on scrubbing technologies and capcity, one of the benefits to this approach is it helps instrument attacks in way that sometimes allows defenders
While paying for protection may seem like a jagged pill to swallow, MaaS providers are arguably our most advanced line of defense.  
In short, the threats are such that is pays to pay for protection.
For example, one of the benefits to this approach is it helps to instrument attacks in way that sometimes allows defenders
(information security teams, incident response teams, etc.) to create profiles that can attribute attacks to specific ``families'' of malcode, and occasionally even attribute
attacks to the actor(s) responsible.
This is especially evident when one considers that large mitigation providers see a broader cross-section of attack traffic, and this enables deeper analysis.
Of course, is that a benefit of MaaS providers' positions, or an indication that our defenses are in need of basic research to overcome the inherent asymmetry 
between attackers and victims?

%% file: remediation.tex
\section{Searching for Remediation}\label{sec:remediation}

{\bf Fundamentals of the problem}
% Nothing has fundamentally changed
In the last 20
% + 
years of fighting DDoS, we have learned a lot, and a lot of insightful systems have been built to counter DDoS attacks.
However, we have not \emph{fundamentally} advanced our \emph{protections}.  
In that time, we have greatly increased the bandwidth of our networks, but that has also benefited our attackers.  
It has actually benefited them more because 
for every remediation instance that has more bandwidth (e.g. scrubbing centers, Internet service instances, redundant sites, etc.),
%%
%% LZ: do you mean "for every remediation instance that deploys more resources"?  e.g. DNS just puts out lots more replicates.
%%
so too does every attacking bot (of which there are more).
In addition, Moore's Law has also brought more abundant, cheap, powerful, and (unfortunately) compromisable end devices on to the Internet at a rate that meets (and often outpaces) our remediation infrastructure.
% Volumetric DDoS uses aggregate power (bandwidth, abundance of compromised/compromisable devices, processing, etc.) to target victims
Even scrubbing centers cannot expect to keep up with the growing edge-capacity of increasingly well provisioned compromised hosts (bots).
This is what frames the asymmetry and ultimate impedance mismatch of DDoS: there are large numbers of attack sources that can arbitrarily send attack traffic, and which 
are being triaged by central remediation infrastructures.
What's more, attack sources are (almost by necessity) far from where we remediate their aggregate attacks.  This leads to congested transit-links, which in turn leads to unobserved (and unobservable) disruption in the network, as attack traffic accumulates on its way to victims.

{\bf What would help: }
With the many types of DDoS, the multiple TTPs, the diverse topologies of routed infrastructure, and more, it is easy to classify the problem-space as problematically complex.
We argue that now is a critical time to embrace a \emph{principled approach} to 
%identifying the architectural problems that enable DDoS attacks.
%% LZ: "enable DDoS" seems a strong accusation. 
%% Maybe "identifying the architectural features that made DDoS attacks easy"?
identifying the architectural features that have made DDoS attacks so relatively easy to launch.
We believe that what is needed are investigations into what fundamentals enable and exacerbate DDoS.
A foundational understanding of this would jumpstart determining what protections are needed, %% which are feasible, 
the possibility (or not) for incrementally deployable solutions, and what operational plans can be effective.
One core observation is that combating the distributed nature of DDoS from relatively centralized vantage points misaligns many core aspects of the nature of DDoS, and enabling remediations at the edge (where attacking nodes reside) seems to be an insightful start at addressing one fundamental aspect of the overall problem.  Approaches along these lines seem to have the potential to begin effectively addressing the scaling problems of volumetric attacks. 
% While perhaps necessary, this is, however, not sufficient for attacks that capitalize on application-level semantics.
%However, given that attacks capitalize on application-level semantics, operationally viable solutions which can be effectuated \emph{in the network}, have seemed illusive.
% However, given that attacks capitalize on application-level semantics, operationally viable solutions which can be effectuated \emph{in the network} requires providing the network layer with information about application semantics -- a requirement conflicting today's TCP/IP layered stack.
However, given that attacks capitalize on application-level semantics, operationally viable solutions which can be effectuated \emph{in the network} require providing the network 
layer with information about application semantics -- a requirement which conflicts with today's TCP/IP layered stack.
%% LZ: I added the above sentence, may need some editing

{\bf Existing Approaches: }
% We don’t have tools to do source address validation
% BCP-38/BCP-84
% Not deployable for transit networks
% Not effective unless deployed from edge inward
In considering how to push remediations into the network, there are some operational and standardized approaches being attempted now.  As discussed in Section~\ref{sec:bg}, FlowSpec, RTBH, and dots all 
attempt flavors of pushing semantics into the network to try and bolster a distributed defense.  
Between operational overheads, coarseness of the remediation, collateral damage, and (in the case of dots) still nascent investigations these 
approaches have at best been triage for our DDoS battlefield damage.
In addition to these, to address source address spoofing, two Best Common Practices 
(BCPs) exist to inform network operators how to configure their networks to no allow out-bound, or in-bound, spoofed packets~\cite{bcp38,bcp84}.  While a great deal of attention
has been paid to getting these deployed, there has been relatively little successful effect (as evidenced by the scale of recent reflector attacks~\cite{mirai-botnet, mansfield2016ddos, krebs2016krebsonsecurity}).
A principled inspection of this might suggest that the reason for limited deployment stems from the misalignment of costs and benefits.  That is, 
since the costs of deploying are not aligned with incentives (i.e. those DDoS victims who benefit from deployment are not the 
network providers who have to deploy and pay if there is a misconfiguration).
Indeed, this has been noted in operational communities as well, 
``The \emph{costs} $\dots$ not directly [being] borne by the potential beneficiaries of deploying the solution''~\cite{huston-apricot19}.

% RPKI
% doesn’t address this (allocation/ctrl-plane not transaction/data-plane)
An interesting question is, since source-address spoofing is related to the inter-domain routing system, should it be mitigated by security protections at that layer?  Today,
inter-domain routing security is being addressed by a relatively new set of standards called the Resource Public Key Infrastructure (RPKI)~\cite{lepinski2012infrastructure}.
However, this approach (and its dependent technologies) do not address the data-plane, and focus only on IP address allocation and potentially some aspects of BGP's
control-plane (which has no relationship to source-address spoofing).

Without a fundamental/principled approach, attack vectors require case-by-case remediation techniques at the application-level.  
What this means is that after each application is discovered to have a DDoS attack vector, 
it and its maintainers must retread an increasingly common path taken by exploited applications: they must create remediation and detection techniques, 
then undergo the process of promoting operational deployment of protections, and then 
sometimes promote associated network configuration changes.  
Examples of applications fending for themselves are DNS' Response Rate 
Limiting (RRL)~\cite{Vixie:2014:RS:2578508.2578510}, the advice to disable NTP's monlist command~\cite{cloudflare-ntp-amplification-ddos-attack}, etc.
%% Many of the application-level attack vectors capitalize on source-address spoofing.

% TVA – path authorization
Mature proposals from the research literature, such as the Traffic Validation Architecture (TVA)~\cite{yang2008tva} and Pushback~\cite{ioannidis2002pushback}, 
have existed for some time.  Approaches like these
aim to offer distributed remediations by using in-network deployment.  Yet, years after publication, they have 
not gained deployment traction.
% As with BCP-38 and BCP-84, deployment of these (and similar technologies) have not materialized.  
As with BCP-38 and BCP-84, incentives are not aligned with those 
paying costs for deployment.
A more recent proposal, Stellar~\cite{dietzel2018stellar}, embraces many of the approaches in FlowSpec, RTBH, and dots by proposing a new black-holing framework.  
It also attempts to better align 
costs with benefits by focusing on deployments in large IXPs, and in doing so also aims to reduce black-holing's collateral damage by pushing remediations 
closer to the attacking sources.
Other recent work called FITT~\cite{DBLP:journals/corr/abs-1902-09033} has begun investigating if today's Internet architecture itself is in need to 
reevaluation. This work proposes using Named Data Networking (NDN)~\cite{Zhang:2014:NDN:2656877.2656887,afanasyev2018brief} as an incrementally deployable solution 
whose incentive model is designed to align costs with benefits,
to (among other things) combat the DDoS threat.

%% {\bf Candidate: remediation towards the edge}
%% Kill-chain in cybersecurity: identify what links in “the chain” of events can be disrupted to stop a threat before it becomes an attack
%% Source-address validation would ameliorate the largest volumetric DDoSes we’ve seen
%% Distributing remediations to disrupt DDoS at its sources (distributed defenses for a distributed threat)
%% Must be coupled with a deployment path: align costs w/ incentives and/or make deployment tractable at scale
%% FlowSpec
%% A standard that allows course network-layer filters to be propagated between providers
%% Problem: often too course grained and when used, not propagated widely
%% BCP-38/BCP-84
%% Would work very well (shedding spoofs)
%% Problem: not deployable after origin because anyone can transit for anyone else, and there is no authority of who is allowed to source traffic
%% Costs not aligned with benefits: why would an origin pay to deploy without gaining any benefit?
%% IETF DDoS Open Threat Signaling (dots) WG (dots)
%% Tries to allow service providers to communicate threats and bolster attack postures
%% Early work, no vendor support, only at network layer

%% file: disc.tex
\section{Discussion}\label{sec:disc}

% Proposals like TVA and similar would make strong contributions, but they need a viable deployment path and may not even go far enough without application-level inspection
% We (community) must do deeper analysis on the fundamentals of the problem to find new solutions
% For example: DDoS is a distributed phenomenon, do distributed remediations match impedances?
% Operationalizing this continues to be an open challenge, but we still must still assess whether this addresses the fundamental  elements of the DDoS phenomena
% For example, the application-level discrimination needed by modern scrubbing services suggests a misalignment of the e2e principle and countermeasures needed for DDoS
% We need to a fresh look finding a new breed of DDoS defenses

Have we made fundamental enhancements to our DDoS defenses in the last 20 years?  The landscape of cheap, compromisable, bots has only become more fertile to miscreants, and more 
damaging to Internet service operators.  Increases in bandwidth have been shared by Internet services and attacking bots, but have been multiplied by a asymmetric scaling factor 
for compromised nodes (there's just more of them).  Our applications have become more complex and even our security and privacy protections (like TLS, HTTPS, etc.) have
made DDoS harder to mitigate in the network.  We need a principled approach to this problem, basic research on ways to bridge an asymmetric gap, incentive models that align costs with
benefits, and novel insights that today's operators can use.  
As a starting point for discussions, we posit that using the network to mount a distributed defense is the right basic approach, and those defense technologies that undercut the 
network properties that DDoS is built on, \emph{and} reward early adopters (economically, qualitatively, or in other palpable ways) are going to be key to changing the tide of our 
war on DDoS.
However, with trends like TTPs increasingly moving to the application-level and the near ubiquity of end-to-end encryption, can we expect the network 
and transport layer semantics to be expressive enough to combat DDoS by themselves?
This is a call to action: the research community is our best hope and best qualified to take up this call.

%% file: acks.tex
%% \begin{acks}
%%	The discussions leading to this editorial were initiated during
%%	Dagstuhl Seminar 15102 on
%%	\emph{Secure Routing for Future Communication Networks},
%%	and we thank all participants for their contributions.
%% \end{acks}